\documentstyle[preprint,prb,aps]{revtex}
\begin{document}
\draft

\title{High -- pressure Raman study of L-alanine crystal}
\author{A. M. R. Teixeira, P. T. C. Freire,$^\dag$ A. J. D. Moreno,$^\ast$
J. M. Sasaki, A. P. Ayala, J. Mendes Filho, F. E. A. Melo}

\address{Departamento de F\'{\i}sica , Universidade Federal do Cear\'a,
Caixa Postal 6030\\
Campus do Pici, 60451-970 Fortaleza, Cear\'a, Brazil}
\maketitle

\begin{abstract}

Pressure-dependent Raman scattering studies in the range 0.0 -- 32
kbar were carried out in L-alanine in order to investigate its
external mode phonon spectra in relation to the phase transitions
in the crystal. A careful analysis of the spectra shows that the
low-energy Raman modes exhibit variation both in frequency and in
intensity and between 26 and 28 kbar it is observed a splitting of
a external mode, indicating that the D$_2$ normal phase undergoes
a transition. Pressure coefficients for external modes are also
given.

\end{abstract}
\vskip1truecm

\pacs{Keywords: A. organic crystals, D. phase transitions, E.
inelastic light scattering.}

\bigskip

\centerline{\bf I. INTRODUCTION}
\bigskip

In recent times one has witnessed a growing interest in the study
of amino acid crystals. This interest has been stimulated by the
perspective of understanding a system where the hydrogen bond
plays a fundamental role and, as a consequence of this
understanding, a better knowledge of some important biological
molecules, e.g. proteins, can be obtained. L-alanine,
NH$_2$CHCH$_3$COOH, is one of 20 amino acid serving as building
blocks for the proteins of living beings. In the crystalline form
L-alanine crystal has been subjected to extensive investigations
by many authors using various experimental techniques. X-ray
diffraction studies revealed the crystal structure and the
position of the heavy atoms of the amino acid.\cite{si}  Neutron
diffraction studies confirmed the results of Ref. 1 and determined
the hydrogen atoms positions.\cite{jo} A number of studies have
been carried out on the vibrational spectra of L-alanine crystal
focusing on the assignment of modes and the effects of various
intermolecular potentials on the crystal vibration \cite{ad,ma},
the effect of temperature on the crystal structure \cite{wa}, the
NH$_3$-torsion temperature dependence \cite{fo} and the dynamic
localization of vibrational energy \cite{mi}. Some studies have
been made on the vibrational spectra of L-alanine
crystal.\cite{ad,ma,wa,fo,mi} Thermal conductivity\cite{kw} and
phonon echo\cite{le} in L-alanine were also performed. NMR
experiments and spin-lattice relaxation time T$_1$ have suggested
that the crystal undergoes a phase transition at 178 K.\cite{ja}
However, Raman spectra do not confirm the existence of this
transition.\cite{wa,mi} It was also suggested that L-alanine
crystals should undergo a phase transition induced by the
application of pressure, \cite{le} but no experimental
investigation was made up to now.

It is well established that the use of hydrostatic pressure can
diminish the interatomic and the intermolecular distances of ions
and molecules in the crystal structure producing, eventually, a
change of structure. Such effects were detected recently by Raman
scattering experiments performed on monohydrated L-asparagine
crystal, \cite{mo} (and confirmed by X-ray diffraction synchrotron
measurements.\cite{sa}), where three different phase transitions
were observed for pressures up to 2.0 GPa.

In this letter we investigate the influence of high hydrostatic
pressure on a L-alanine crystal observing the low-frequency region
of the Raman spectra. By exploring the fact that the modes of the
low-frequency region gives important information about the
structure's stability we performed Raman measurements in the range
of pressure 0.0 -- 32 kbar. We also give pressure coefficients for
the external modes, fundamental data to determine Gr\"uneisen
parameter.

\bigskip

\centerline{\bf II. EXPERIMENTAL}
\bigskip

High-pressure Raman experiments at room temperature were performed
on a small piece of L-alanine  sample pressed in a standard
Diamond Anvil Cell. As hydrostatic medium we have used a 4:1
methanol:ethanol mixture. The pressure was calibrated using the
shifts of ruby lines. The pressure calibration is expected to be
accurate by $\pm$ 0.3 kbar. The Raman spectra were excited with a
514.5 nm line of an Argon ion laser working at 30 mW in the
backscattering geometry. The laser beam was focused on the sample
surface using a lens with f = 20.5 mm. To ensure focusing of laser
on the sample when Raman spectra were recorded or on the ruby chip
when pressure was calibrated a image of the hole in the gasket,
the pressure compartment, was recorded by a CCD camera. The
backscattering light was analyzed in a triplemate Jobin Yvon
spectrometer (T64000) equipped with a N$_2$ - cooled CCD system.
The frequency of the Raman bands are expected to be accurate by
$\pm$ 2 cm$^{-1}$.

\bigskip
\centerline{\bf III. RESULTS AND DISCUSSIONS}
\bigskip

The L-alanine crystal has an orthorhombic symmetry with four
molecules per unit cell with a space group P2$_1$2$_1$2$_1$
(D$_2^4$) and cell parameters a = 6.023, b = 12.343 and c = 5.784
$\AA$. The molecule is found in the zwitterion form and all three
hydrogen atoms of the ammonium group form hydrogen bonds with
three different neighboring molecules. In fact, the main link of
molecules in the crystal structure occurs along the c-axis, where
it is observed a chain of hydrogen bonds. Also, two other hydrogen
bonds bind these chains together producing a three dimensional
network. \cite{si} Such link mechanism is common to most amino
acid crystals as L-threonine \cite{sh,silva} and monohydrated
L-asparagine \cite{ram}.

It is well established that the intramolecular modes are in the
high frequency region of the Raman spectra of molecular crystals
while the external phonon modes, representing essentially the
translations and the rotations of rigid molecules, are in the low
frequency region. The pressure effect on the crystal vibrations
are mainly of two types: (a) slight modifications on the energy of
modes in the high frequency region; (b) large shifts of the energy
modes in the low frequency region. These modifications are due to
the fact that the pressure should greatly decrease the
intermolecular distances while the interatomic separations of each
molecule are only slightly decreased. Eventually, if the structure
is greatly modified by the external agent (pressure) via hard
changes in the intermolecular distances, it will adapt itself to a
new configuration and the crystal symmetry will be changed.
Associated to these changes it will be observed drastic
modifications in the external modes region of the Raman spectra as
have been already observed in another amino acid crystal.
\cite{mo}

In Fig. 1 we show Raman spectra of L-alanine crystal for selected
pressures. The spectra are unpolarized but the laser beam is
reaching the sample surface along the c-axis. All pressure values
are given in kbar units. The spectrum taken at 0.7 kbar has a
profile similar to the spectrum recorded at atmospheric pressure,
except for little differences in the relative intensities. In this
spectrum we labeled the external mode peaks as A, B, C, D, E and
F, in order to discuss the effect of pressure in a most clear way.
The peak A has an initial frequency of 41 cm$^{-1}$, while the
peaks B, C, D, E and F have frequencies at P = 0.0 kbar equal to
48, 75, 105, 113 and 138 cm$^{-1}$, respectively. The 41 and 48
cm$^{-1}$ modes have been assigned as $w$-axis libration, in a
picture where $u, v$ and $w$ are perpendicular axes and,
additionally, the $v$ is along the long molecular axis, $w$ is
perpendicular to the plane of the molecule and $u$ is nearly
parallel to the crystallographic c-axis. \cite{loh} It was
observed through temperature studies an instability in the 48
cm$^{-1}$ mode associated to localization of vibrational energy in
the 41 cm$^{-1}$ mode. \cite{mi} Although the localization of
vibrational energy should not be related to the pressure effect in
a direct way, it is worth to mention that as pressure is
increased, the peak with frequency 41 cm$^{-1}$ gains intensity,
similar to the temperature effect. In fact we observe that at 0.7
kbar the intensity of the bands of 41 and 48 cm$^{-1}$ modes are
almost the same and as the pressure increases, the difference
between the intensity of the two modes increases considerably.
However the main effect of pressure on the Raman spectra of
L-alanine crystal is observed in another spectral region as
discussed in a forthcoming paragraph.

The dependence with pressure of all L-alanine low frequency bands,
can be seen by analysis of the plots of frequency ($\omega$)
versus pressure (P) given in Fig. 2. In the plots the circles
represent data taken while compressing the L-alanine crystal up to
32.1 kbar and the solid lines are the least square fitting of the
data to the function:
\begin{equation}
\omega = \omega_0 + \alpha P \label{eq1}
\end{equation}
A linear behavior was observed for all the $\omega$ vs. P plots of
low frequency modes. The values found for the intercept,
$\omega_0$, the linear pressure coefficient, $\alpha$, and the
experimental frequency of the mode at P = 0 kbar,
$\omega_{0(exp.)}$, are listed in Table I. It is important to
state that for peak A, the linear fitting is a good one for data
from 0 to 18 kbar (the $\omega_0$ value for peak A in Table I
corresponds to this first linear fitting). Also, it is observed
that from 20 to 32 kbar a linear fitting adjust very well to
experimental data of peak A, but with a different dependence of
frequency with pressure. This would mean that a slow modification
of the L-alanine structure is starting to occur in a pressure
value of about 20 kbar.

This modification, in fact, will be completed at a pressure of
about 28.4 kbar. In Fig. 1 the spectrum taken at 26.8 kbar shows
that peak E is just a single band and when pressure reaches 28.4
kbar a splitting of the peak is verified. The spectrum of 28.4
kbar in Fig. 1 shows this new peak arising from the splitting of
peak E, and it is marked as G. In Fig. 2 the plots of $\omega$
versus P also illustrates well the splitting of peak E above 28
kbar. Such a splitting is accounted for as a result of change of
the lattice symmetry caused by a modification of the elementary
cell. In other words, the number of external modes increases when
the L-alanine crystal is compressed above 28 kbar and a phase
transition to a new structure takes place.

It may be worthwhile to mention and to compare the high pressure
L-alanine crystal study with that did in another amino acid
crystal, monohydrated L-asparagine,
NH$_{2}$CO(CH$_2$)CH(NH$_2$)COOH $\bullet$ H$_2$O \cite{mo,sa}.
Monohydrated L-asparagine crystal undergoes phase transitions at
1.0, 6.0 and 13 kbar, pressure values below the pressure value of
the phase transition in L-alanine. The phase transition observed
on monohydrated L-asparagine was ascribed to be a consequence of
shortening of the length of the hydrogen (H) bonds in the crystal
structure. Indeed, there are seven hydrogen bonds involving all
the hydrogen atoms attached to the two nitrogen atoms and to
oxygen of water molecule. These H bonds are fundamental to link
together adjacent chains of molecules, including the bonds
originated from water molecule. On the other hand, the
zwitterionic L-alanine does not involve any water molecule. It is
likely that the greater stability under high pressure of L-alanine
compared with the other amino acid crystal should be originated
from the fact that the little glue group of monohydrated
L-asparagine, the water molecule, is absent in the material we
discussed in this letter. Central to our understanding of the
difference of the pressure effect on the two amino acid crystals
is the possibility of localization of the atoms participating of
the hydrogen bonds. Of course, a more precise description of the
shortening of the molecular links and the right mechanism of the
phase transition in L-alanine crystal will be achieved by the
localization of the heavy atoms of the molecule by an x-ray
diffraction study. In summary, comparing the monohydrated
L-asparagine crystal with the L-alanine crystal a more stable
structure under pressure is achieved for the last material.

\bigskip
\centerline{\bf IV. CONCLUSIONS}
\bigskip

The results of our pressure Raman light scattering studies
performed for the L-alanine crystal have led to the following
conclusion: for pressures up to 32 kbar the crystal undergoes a
phase transition between 26.8 and 28.4 kbar. The transition
observed in our work was identified by a drastic change in the
Raman spectra, i.e., an increasing in the number of observed Raman
peaks in the external mode region. The pressure coefficients for
external mode was also given.

It is clear that deductions concerning structural changes based on
the spectra reported here, although very clear, is not complete.
X-ray diffraction measurements over this pressure range, both on
the room pressure phase and on the new phase would be of great
value for determining the new space group and understanding the
mechanisms of the changes revealed in this study.

\acknowledgments

We acknowledge the financial support from the FUNCAP, CAPES and
FINEP, Brazilian agencies. We are indebted to Prof. O. Pilla for
discussions and to Prof. M.P. Almeida for a reading of the
manuscript.

\begin{table}
\label{tableI} \caption{Raman frequencies of L-alanine crystal at
P = 0 kbar,  $\omega _{0(exp.)}$, $\omega _0$, and $\alpha$ value
for a fitting of the type  $\omega$ = $\omega _0$ + $\alpha$ P.
The letters represent the peaks as stated in Fig. 1.}
\begin{center}
\begin{tabular}{cccc}
Peak &  $\omega_{0(exp.)}$ (cm$^{-1}$) &  $\omega_0$ (cm$^{-1}$) &
$\alpha$ (cm$^{-1}$ kbar $^{-1}$) \\ \hline
 A & 40 & 41.7 & 0.07 \\
 B & 48 & 48.0 & 0.45 \\
 C & 75 & 76.3 & -0.03 \\
 D & 105 & 105.3 & 0.38 \\
 E & 113 & 115.3 & 0.83 \\
 F & 139 & 141.5 & 1.24

\end{tabular}
\end{center}

\end{table}

\begin{figure}
\label{Fig.1} \caption{Raman spectra of L-alanine crystal at
various pressure values. The numbers indicate the pressure in
units of kbar.}
\end{figure}

\begin{figure}
\label{Fig.2} \caption{Dependence of frequency of the external
modes with pressure for L-alanine crystal. The circles are data
for increasing pressure. The solid lines are fittings to the
experimental data whose coefficients are given in Table 1.}
\end{figure}


\begin{references}

\bibitem[\dag]{ad1}  Corresponding author; E-mail: tarso@fisica.ufc.br

\bibitem[\ast]{ad2} Permanent adress: Departamento de F\'{\i}sica,
Universidade Federal do Maranh\~ao

\bibitem{si} H. J. Simpson and R. E. Marsh, Acta Cryst. {\bf 20},
550 (1966).

\bibitem{jo} P. G. Jonsson and A. Kvick, Acta Cryst.
{\bf B28}, 1827 (1972).

\bibitem{ad} R. Adamowicz and E. Fishman, Spectrochim. Acta {\bf 28A},
889 (1972).

\bibitem{ma} K. Machida, A. Kagayama, Y. Saito, and T. Uno, Spectrochim.
Acta {\bf 34A}, 909 (1978).

\bibitem{wa} C. H. Wang and R. D. Storms, J. Chem. Phys. {\bf 55},
3291 (1971).

\bibitem{fo} S. Forss,  J. Raman Spectrosc. {\bf 12}, 266 (1982).

\bibitem{mi} A. Migliori, P. Maxton, A. M. Glogston, E. Zirngiebel
and M. Lowe, Phys. Rev. B {\bf 38}, 13464 (1998).

\bibitem{kw} R. S. Kwok, P. Maxton, and Migliori, Solid State
Commun. {\bf 74}, 1193 (1990).

\bibitem{le} V. V. Lemanov and R. K. Harris, Phys. Solid State
{\bf 40}, 1922 (1998).

\bibitem{ja} P. Jackson and R. K. Harris, J. Chem. Soc. Faraday
Trans. {\bf 91}, 805 (1995).

\bibitem{mo} A. J. D. Moreno, P. T. C. Freire, F. E. A. Melo,
M. A. Ara\'ujo Silva, I. Guedes, and J. Mendes Filho,  Solid State
Commun. {\bf 103}, 655 (1997).

\bibitem{sa} J. M. Sasaki, {\it et. al.} to appear in Proceeding
of the XVII International Conference on High Pressure Science and
Technology.

\bibitem{sh} D. P. Shoemaker, J. Donohue, V. Schomaker and R. B. Corey,
J. Am. Chem. Soc. {\bf 72}, 2328 (1950).

\bibitem{silva} B. L. Silva, P. T. C. Freire, F. E. A. Melo, I. Guedes,
M. A. Ara\'ujo Silva, J. Mendes Filho, A. J. D. Moreno, Braz. J.
Phys. {\bf 28}, 19 (1998).

\bibitem{ram} M. Ramanadham, S. K. Sikka and R. Chidambaram, Acta Cryst. B
{\bf 28}, 3000 (1972).

\bibitem{loh} E. Loh, J. Chem. Phys. {\bf 63}, 3192 (1975).


\end{references}
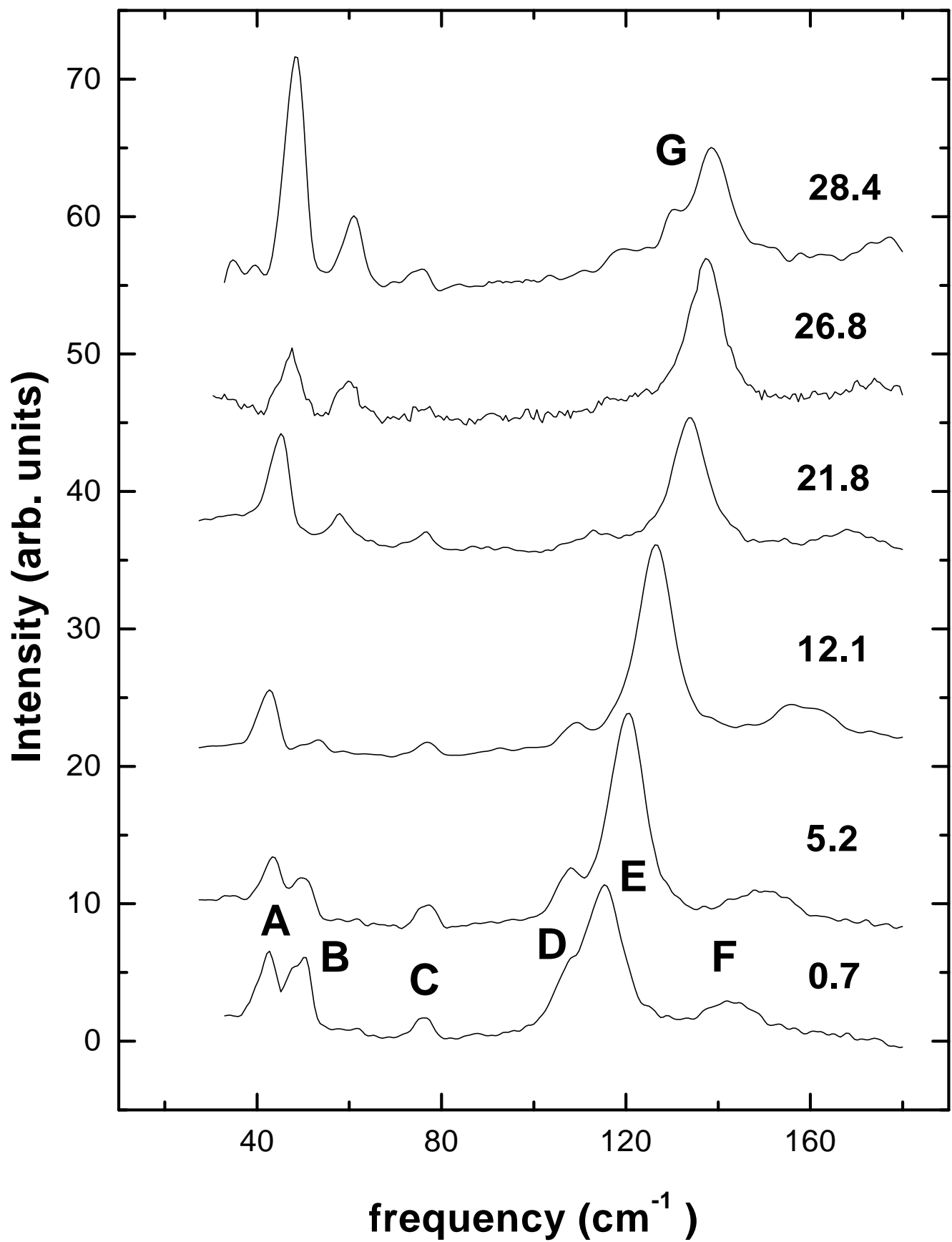
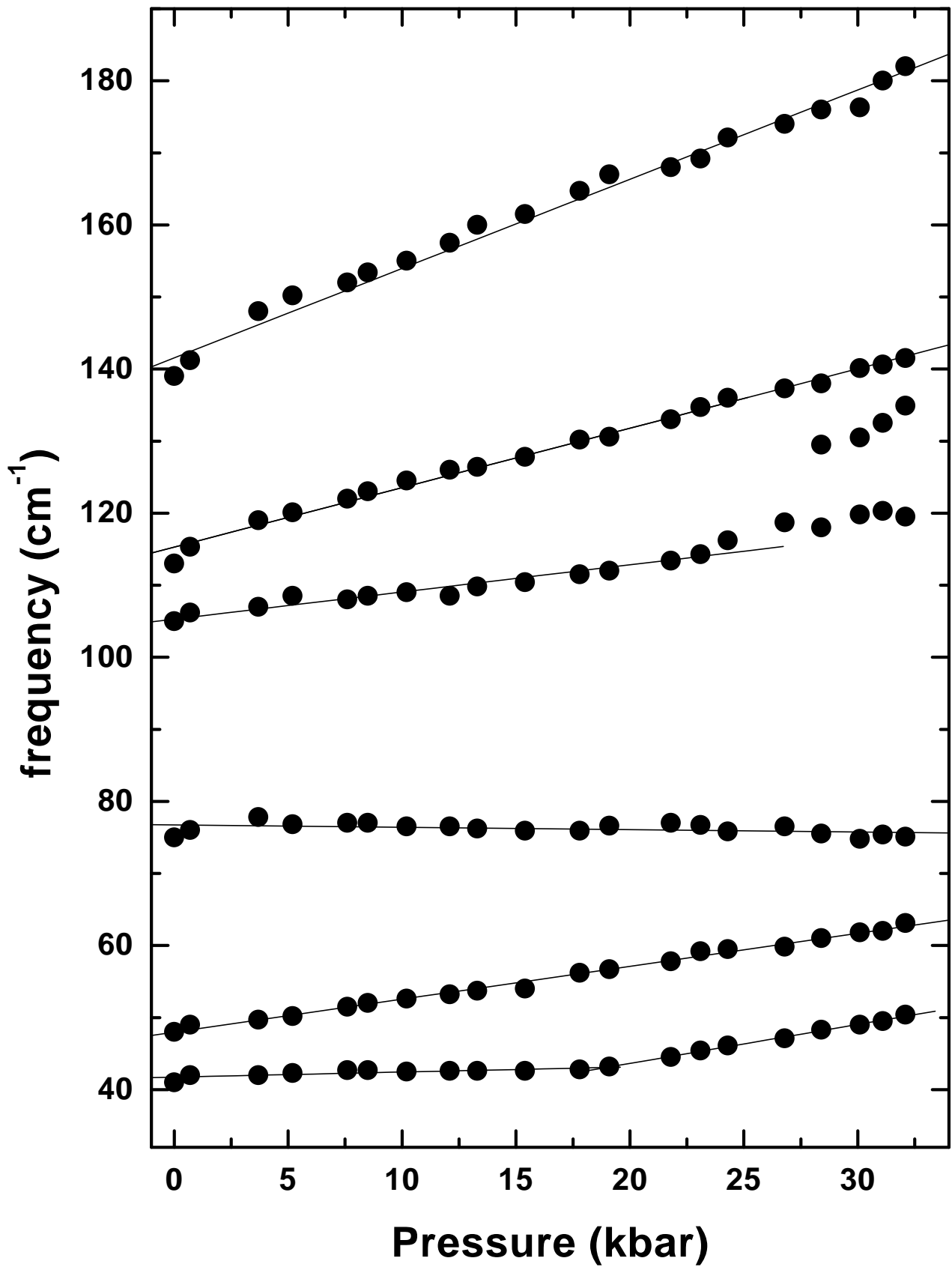
\end{document}